\documentclass[lettersize,journal]{IEEEtran}
\usepackage{amsmath,amsfonts}
\usepackage{algorithmic}
\usepackage{makecell}
\usepackage{array}
\usepackage[caption=false,font=normalsize,labelfont=sf,textfont=sf]{subfig}
\usepackage{textcomp}
\usepackage{stfloats}
\usepackage{amssymb}
\usepackage{url}
\usepackage{verbatim}
\usepackage{graphicx}
\usepackage{booktabs}
\usepackage{dsfont}
\usepackage[T1]{fontenc}
\def\BibTeX{{\rm B\kern-.05em{\sc i\kern-.025em b}\kern-.08em
    T\kern-.1667em\lower.7ex\hbox{E}\kern-.125emX}}
\usepackage{balance}
\usepackage{cite}
\begin{document}
\title{Syndrome-as-Header: A Quantum Label-Switching Architecture via Uncorrectable Error Injection}
\author{IlKwon Sohn, Ju-Bong Kim, Kwangil Bae, Wooyeong Song, Wonhyuk Lee
\thanks{%Manuscript received July 15, 2024;
This research was supported by Korea Institute of Science and Technology Information(KISTI)(K26L1M3C5). This research was supported by the National Research Council of Science \& Technology(NST) grant by the Korea government (MSIT) (No. CAP22053-000)

IlKwon Sohn, Ju-Bong Kim, Kwangil Bae, Wooyeong Song, and Wonhyuk Lee are with the Quantum Network Research Center, Korea Institute of Science and Technology Information, Daejeon 34141, Republic of Korea (e-mail: d2estiny@kisti.re.kr).}}
%\markboth{Journal of \LaTeX\ Class Files,~Vol.~18, No.~9, September~2020}%
%{How to Use the IEEEtran \LaTeX \ Templates}

\maketitle

\begin{abstract}
Scaling quantum networks beyond point-to-point links requires packet forwarding that can tolerate control-plane timing uncertainty.
Existing optical-burst switching (OBS) and quantum-wrapper-based packet switching (QW) rely on external classical headers whose jitter-prone processing must remain aligned with in-flight quantum payloads, creating guard-window or fiber-delay-line budget failures.
This paper proposes Syndrome-as-Header (SAH), a quantum label-switching architecture that embeds routing labels into the syndrome structure of an encoded payload.
The proposed scheme uses Uncorrectable Error Injection (UEI) to map flow labels to reference syndromes and builds a syndrome-space header codebook whose decoding regions remain distinguishable under correctable channel-induced syndrome deviations.
Core routers extract the syndrome header, suppress the correctable residual channel-error component, and swap labels without measuring or decoding the logical payload.
SAH supports FAST forwarding and VERIFICATION mode, the latter providing end-to-end consistency checking while retaining swappable labels.
In NSFNet timing benchmarks with common per-hop quantum error correction (QEC), SAH eliminates the external-header/payload alignment condition, so timing uncertainty appears as memory residence and delivered latency rather than alignment-budget drops.
With a $T_2$-based memory-residence penalty, the advantage depends on router memory coherence; for sufficiently long coherence, SAH maintains higher acceptance and throughput than OBS+QEC and QW+QEC under high jitter.
\end{abstract}


\section{Introduction}
\IEEEPARstart{T}{he} transition from static circuit-switched architectures to dynamic packet-switched architectures~\cite{fo13} is essential for scaling quantum networks to support multi-party applications~\cite{ki08, we18, ca19}.
Such quantum networks will carry heterogeneous traffic with diverse payload sizes and coding overheads, requiring forwarding mechanisms that are not constrained by a fixed logical-payload format~\cite{rv14, di22, yo24, da19}.

Dynamic switching, however, is constrained by limited buffering of in-flight quantum payloads and by jitter-prone control latency~\cite{di22, ca202}: when forwarding depends on a separate timing-critical classical header, synchronizing the switch decision with payload arrival requires offset/guard times or physical delay elements~\cite{di22, yo24}, which becomes a bottleneck under control-plane timing uncertainty.

Existing quantum packet-switching approaches typically follow either offset-based optical burst switching (OBS)~\cite{qi99, di22, ma07} or quantum-wrapper-based packet switching (QW)~\cite{yo24}.
OBS sends a classical control header ahead of the quantum payload, making forwarding sensitive to offset fluctuations and guard-window violations.
QW co-propagates the header and payload within the same datagram and uses fiber delay lines during header processing, which relaxes offset scheduling at the cost of additional hardware and a fidelity--timing trade-off~\cite{yo24}.
Thus, both paradigms keep forwarding tied to an external classical header whose timing must remain aligned with the quantum payload.

As the Quantum Internet moves toward tighter interconnection among quantum computers~\cite{we18}, core nodes are expected to combine quantum memories~\cite{bh20, az23} with quantum error correction (QEC) primitives~\cite{mu14, ni23} to preserve encoded states across routing operations.
In such a QEC-enabled core, syndrome extraction and recovery become inherent per-hop operations on encoded payload blocks~\cite{mu16, sh95, gt97, st96}.

Adding QEC processing to OBS or QW can protect the encoded payload against correctable physical errors, but it leaves the forwarding information outside the encoded block.
Therefore, the timing dependence identified above remains even in QEC-equipped OBS+QEC and QW+QEC baselines.
This observation motivates a QEC-native forwarding architecture in which the syndrome structure itself is used as a forwarding resource.

The proposed Syndrome-as-Header (SAH) architecture realizes this direction by using the syndrome structure of an existing quantum error correction code (QECC) to represent in-band routing labels, rather than attaching or co-propagating a separate timing-critical classical header.
SAH employs Uncorrectable Error Injection (UEI)~\cite{so25} to embed routing labels directly into the syndrome space of an encoded quantum payload, consistent with prior work showing that controlled syndrome outcomes can carry classical information~\cite{ch20}.
Each compact flow label is mapped to a UEI-induced admissible reference syndrome through a syndrome-space header codebook, whose decoding regions remain disjoint under correctable channel-induced syndrome deviations.
SAH thus relies only on per-hop memory residence at routers rather than long-term quantum storage.

Because the forwarding label is embedded in the syndrome domain of the encoded block, routers can recover labels, suppress correctable per-hop channel-error deviations, and swap labels via syndrome processing without logical decoding.
SAH supports FAST mode for autonomous label switching without real-time classical coordination and VERIFICATION mode for egress-side checking of the expected flow-specific verification component before payload release~\cite{so25}.

The contributions of this paper are threefold.
First, we introduce SAH as a QEC-native, syndrome-domain switching abstraction that removes strict external-header/payload timing alignment from quantum packet forwarding.
Second, we formulate a syndrome-space header codebook and an autonomous UEI-based label-swap protocol for encoded payload blocks.
Third, we extend the proposed scheme with an end-to-end verification mode and evaluate it against QEC-equipped OBS+QEC and QW+QEC baselines under control-plane jitter, including a memory-residence penalty parameterized by the physical memory coherence time $T_2$ of router memories~\cite{bh20,az23}.

\section{Preliminaries}
\subsection{Quantum error correction and syndromes}
QECCs~\cite{sh95, st96, gt97} encode $k$ logical qubits into $n$ physical qubits so that physical errors can be detected and corrected through syndrome measurements.
In the stabilizer formalism~\cite{gt97}, an $[[n,k,d]]$ stabilizer code is specified by a stabilizer group $\mathcal{S}\subset\mathcal{P}_n$, generated by $\{g_i\}_{i=1}^{n-k}$, where $\mathcal{P}_n$ is the $n$-qubit Pauli group.
Throughout the paper, Pauli operators are identified up to global phases in $\{\pm1,\pm i\}$, since these phases do not affect commutation relations with stabilizer generators or measurement statistics.
The code space is the simultaneous $+1$ eigenspace of all elements in $\mathcal{S}$.
We assume a fixed public recovery rule that maps each measured syndrome to a Pauli recovery representative.

For a Pauli error $E\in\mathcal{P}_n$, the syndrome bit $s_i(E)$ equals $1$ if and only if $E$ anticommutes with $g_i$, and equals $0$ otherwise.
We write the syndrome map as
\begin{equation}
s(E) = (s_1(E),\ldots,s_{n-k}(E))\in\mathbb{F}_2^{n-k},
\end{equation}
where $\mathbb{F}_2 = \{0,1\}$ denotes the binary field with arithmetic modulo $2$.
For Pauli operators $E_1,E_2\in\mathcal{P}_n$, the syndrome map is additive over $\mathbb{F}_2$:
\begin{equation}
\label{eq:syn_add}
s(E_1E_2) = s(E_1)\oplus s(E_2).
\end{equation}
This additivity is the key property used later to separate UEI-induced reference syndromes from channel-induced syndrome deviations.

For an $[[n,k,d]]$ stabilizer code, up to $t=\lfloor(d-1)/2\rfloor$ Pauli errors can be corrected.
We define the correctable Pauli-error set as
\begin{equation}
\mathcal{E}_{\mathrm{corr}} := \{E\in\mathcal{P}_n:\mathrm{wt}(E)\le t\}.
\end{equation}
The measured syndrome determines the recovery prescribed by the public recovery rule, not necessarily a unique physical error.
For degenerate codes, distinct correctable errors may share the same syndrome while being corrected by the same recovery action on the code space.

We fix a public recovery rule $C:\mathbb{F}_2^{n-k}\to\mathcal{P}_n$ such that, for every $E\in\mathcal{E}_{\mathrm{corr}}$,
\begin{equation}
C(s(E))E\in\mathcal{S}.
\end{equation}
Thus, the combined operation acts as the identity on the code space.
An intermediate node can therefore apply $C(s)$ to suppress correctable channel-induced errors without measuring the logical information or fully decoding the payload.

\subsection{Uncorrectable error injection mechanism}
\label{UEI}
UEI was originally introduced in~\cite{so25} as a mechanism for fault-tolerant and secure long-distance quantum communication, where a designated Pauli operator outside the correctable set is used to control syndrome outcomes.
Whereas channel-induced Pauli errors are assumed to be correctable when they lie in $\mathcal{E}_{\mathrm{corr}}$, UEI intentionally uses a Pauli operator outside this set as an in-band marker.
We denote the UEI marker operator by $E_{\mathrm{un}}$ and a channel-induced Pauli error by $E_{\mathrm{ch}}$.
Let $\Sigma_{\mathrm{corr}} := \{s(E):E\in\mathcal{E}_{\mathrm{corr}}\}$ denote the set of syndromes induced by correctable Pauli errors.
We choose UEI operators whose induced syndromes satisfy
\begin{equation}
s(E_{\mathrm{un}})\notin\Sigma_{\mathrm{corr}},
\end{equation}
and refer to the resulting syndromes as admissible reference syndromes.
This admissibility condition keeps the intentional UEI marker outside the zero-centered correctable-syndrome region associated with ordinary quantum error correction.

By the additivity of the syndrome map in~\eqref{eq:syn_add}, a correctable channel error $E_{\mathrm{ch}}\in\mathcal{E}_{\mathrm{corr}}$ occurring after the UEI operation produces a measured syndrome
\begin{equation}
s_{\mathrm{meas}} = s(E_{\mathrm{un}})\oplus s(E_{\mathrm{ch}}).
\end{equation}
The residual syndrome deviation
\begin{equation}
\label{eq:res_syn}
\Delta s := s_{\mathrm{meas}}\oplus s(E_{\mathrm{un}}) = s(E_{\mathrm{ch}})
\end{equation}
lies in $\Sigma_{\mathrm{corr}}$, so $C(\Delta s)$ suppresses the channel-induced component while preserving the injected reference component.

\section{System model}
\subsection{Network model and assumptions}
We consider a quantum label-switching network inspired by the Multiprotocol Label Switching (MPLS) principle~\cite{rs01}, with ingress/egress nodes and intermediate core routers.
Forwarding is based on locally significant flow labels rather than global destination addresses.
The architecture consists of a quantum data plane that transports encoded qubit payloads and a classical control plane that pre-distributes label bindings, syndrome-codebook information, and forwarding state, such as label-swap rules~\cite{da19}.
The per-block forwarding information carried by an encoded payload is represented in the syndrome domain.
The classical control plane is assumed to be authenticated and sufficiently reliable for distributing label bindings, syndrome-codebook information, and forwarding state.

Each quantum packet consists of one or more logical blocks encoded using a common $[[n,k,d]]$ stabilizer code whose stabilizer group $\mathcal{S}$ and public recovery rule are fixed and known to all network nodes.
For a multi-block packet, the same packet-level forwarding label is embedded independently in each encoded block, so that syndrome extraction and channel-error correction can be performed blockwise.

We assume memory-assisted forwarding nodes that can store an encoded block during local syndrome extraction, Pauli correction, forwarding-state lookup, and UEI-based label operations.
Such processing may be implemented through a light--matter--light interface, as in memory-assisted and QEC-enabled quantum repeater architectures~\cite{mu14, bh20, az23}.
SAH therefore requires local per-hop memory residence for processing, rather than long-term end-to-end quantum storage.

When physical errors are explicitly considered, each hop may introduce an independent Pauli error $E_{\mathrm{ch}}$ on the transmitted encoded block.
These discrete per-hop Pauli errors are treated separately from decoherence accumulated during local memory residence.
The former is handled through the shared QEC model, whereas the latter is associated with memory-assisted forwarding.

\subsection{Compact syndrome-domain header format}
We use classical link-layer formats such as Ethernet~\cite{ie22} only as a familiar reference for address-oriented header overhead, not as a target data-plane format for quantum packets. 
This comparison motivates a compact syndrome-domain header: embedding conventional address-style fields in the syndrome domain would unnecessarily inflate the syndrome-header budget. 
Because path setup, label binding, and forwarding-state distribution are handled by the separate classical control plane, the in-band header does not need to carry a full destination address or payload-type field. 
SAH therefore retains only the forwarding information required inside the quantum switching core, embedded in each encoded block through its UEI-induced reference syndrome. 
In the benchmarked format, this compact header consists of two fields.

\begin{itemize}
\item \emph{Flow label:} A locally significant identifier that selects a forwarding path and supports label swapping at routers.
\item \emph{Sequence number:} A short per-flow counter used by the egress node to detect limited reordering, loss, or duplication within a bounded window; the 4-bit size is illustrative and can be adjusted with the header budget.
\end{itemize}

\begin{figure}[ht!]
\centering
\includegraphics[width=1\linewidth]{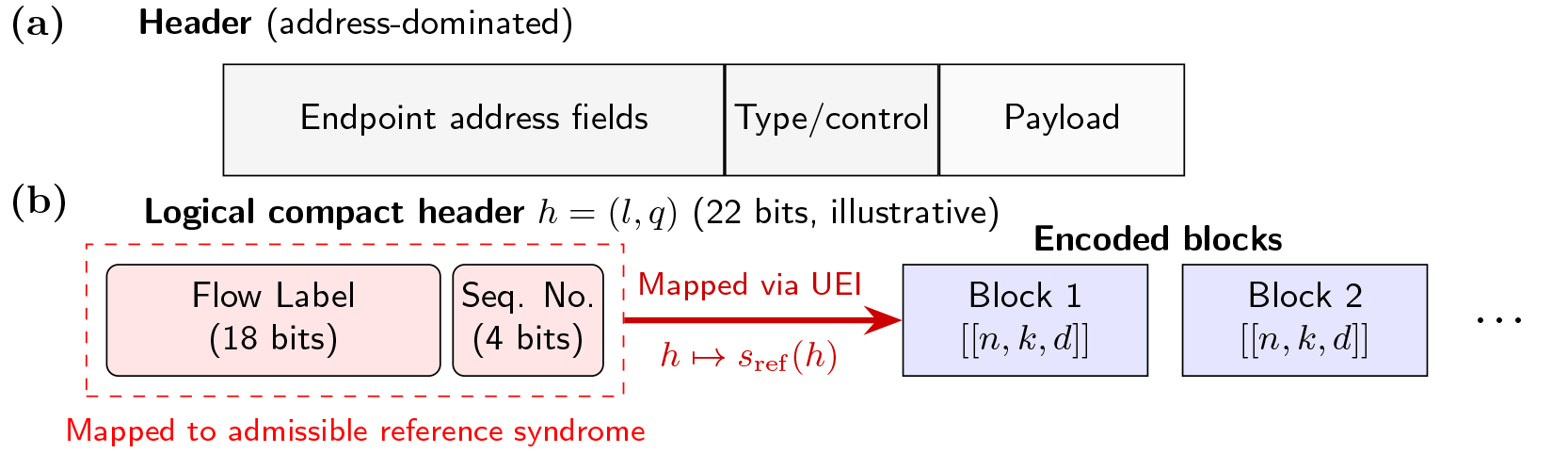}
\caption{Illustrative comparison of forwarding-header abstractions.
(a) An address-dominated classical header carries endpoint address fields, type/control information, and payload.
(b) The proposed logical compact header $h=(\ell,q)$ contains only a flow label and a sequence number, which are mapped via UEI to an admissible reference syndrome $s_{\mathrm{ref}}(h)$ and embedded across encoded blocks.
The comparison highlights compact syndrome-domain forwarding without requiring a classical frame format for quantum packets.}
\label{fig:header}
\end{figure}

The bit widths in Fig.~\ref{fig:header}(b) specify the nominal logical-header fields, whereas the number of simultaneously usable robust headers is determined by the syndrome-space header codebook introduced below.
SAH maps only an active compact-header set to admissible reference syndromes, subject to a disjoint decoding-region condition and a packing constraint on the syndrome space.
Larger label spaces can be supported by larger stabilizer codes, a different active-header allocation, or an extended multi-block header-framing design.
This compact format defines the syndrome-domain forwarding information used by the SAH protocol.

\subsection{Syndrome-space header codebook and control-plane state}
\label{subsec:label_codebook}
Building on the UEI mechanism introduced in Sec.~\ref{UEI}, we specify the static control-plane state and the syndrome-space header codebook required by SAH.
All network nodes use the common stabilizer code, stabilizer-generator ordering, syndrome map, correctable-syndrome set $\Sigma_{\mathrm{corr}}$, and public recovery rule defined above.
These shared conventions allow reference syndromes to be interpreted consistently across the network.

The choice of the underlying $[[n,k,d]]$ stabilizer code and the syndrome-space header codebook is a design parameter determined by the physical channel error rate, the required logical protection level, and the target header space.
The stabilizer code sets the error-correction capability $t$ and the syndrome dimension $n-k$.
The header codebook selects admissible reference syndromes in $\mathbb{F}_2^{n-k}$ outside the zero-centered correctable-syndrome region $\Sigma_{\mathrm{corr}}$, so that intentional header syndromes are separated from ordinary correctable-syndrome deviations.
The header codebook need not be a linear subspace; when algebraic structure is desired, it may be constructed as a coset or shifted version of a classical binary code.

Using the compact header defined above, we write $h=(\ell,q)$ with forwarding label $\ell$ and sequence number $q$.
For router $r$, the forwarding state is represented by a local mapping
\begin{equation}
(p,\ell')=\pi_r(\ell),
\end{equation}
where $p$ is the selected output port and $\ell'$ is the outgoing label associated with the incoming label $\ell$.
The sequence number $q$ is not used for forwarding-state lookup and is preserved across label swaps, so the corresponding outgoing header is $h'=(\ell',q)$.

The classical control plane installs a public header table and per-router forwarding state.
The public header table specifies a finite active compact-header set $\mathcal{H}_{\mathrm{act}}\subseteq\{0,1\}^{b}$ and a deterministic injective mapping
\begin{equation}
f:\mathcal{H}_{\mathrm{act}}\rightarrow\{0,1\}^{n-k},
\end{equation}
whose image is restricted to admissible reference syndromes and chosen to satisfy the disjoint decoding-region condition below.
For each $h\in\mathcal{H}_{\mathrm{act}}$, we define
\begin{equation}
s_{\mathrm{ref}}(h):=f(h).
\end{equation}
The same table also fixes a logical-neutral UEI labeling operator $E_{\mathrm{lab}}(h)$ satisfying $s(E_{\mathrm{lab}}(h))=s_{\mathrm{ref}}(h)$.
To make the representative choice explicit, let $N(\mathcal{S})$ denote the normalizer of the stabilizer group $\mathcal{S}$, and fix a canonical pure-error representative $T(s)$ for each syndrome $s$, with $s(T(s))=s$.
For each active header $h$, the public operator table selects a representative satisfying
\begin{equation}
E_{\mathrm{lab}}(h) \in T(s_{\mathrm{ref}}(h))\mathcal{S},
\end{equation}
or equivalently,
\begin{equation}
T(s_{\mathrm{ref}}(h))^\dagger E_{\mathrm{lab}}(h) \in \mathcal{S}.
\end{equation}
Thus, after removing the prescribed pure-error representative, the remaining zero-syndrome operator is a stabilizer rather than a nontrivial logical Pauli operator in $N(\mathcal{S})\setminus \mathcal{S}$.
This restriction ensures that header injection changes the syndrome sector of the encoded block but does not apply an unintended logical operation to the payload.

\begin{figure*}[t]
\centering
\includegraphics[width=\linewidth]{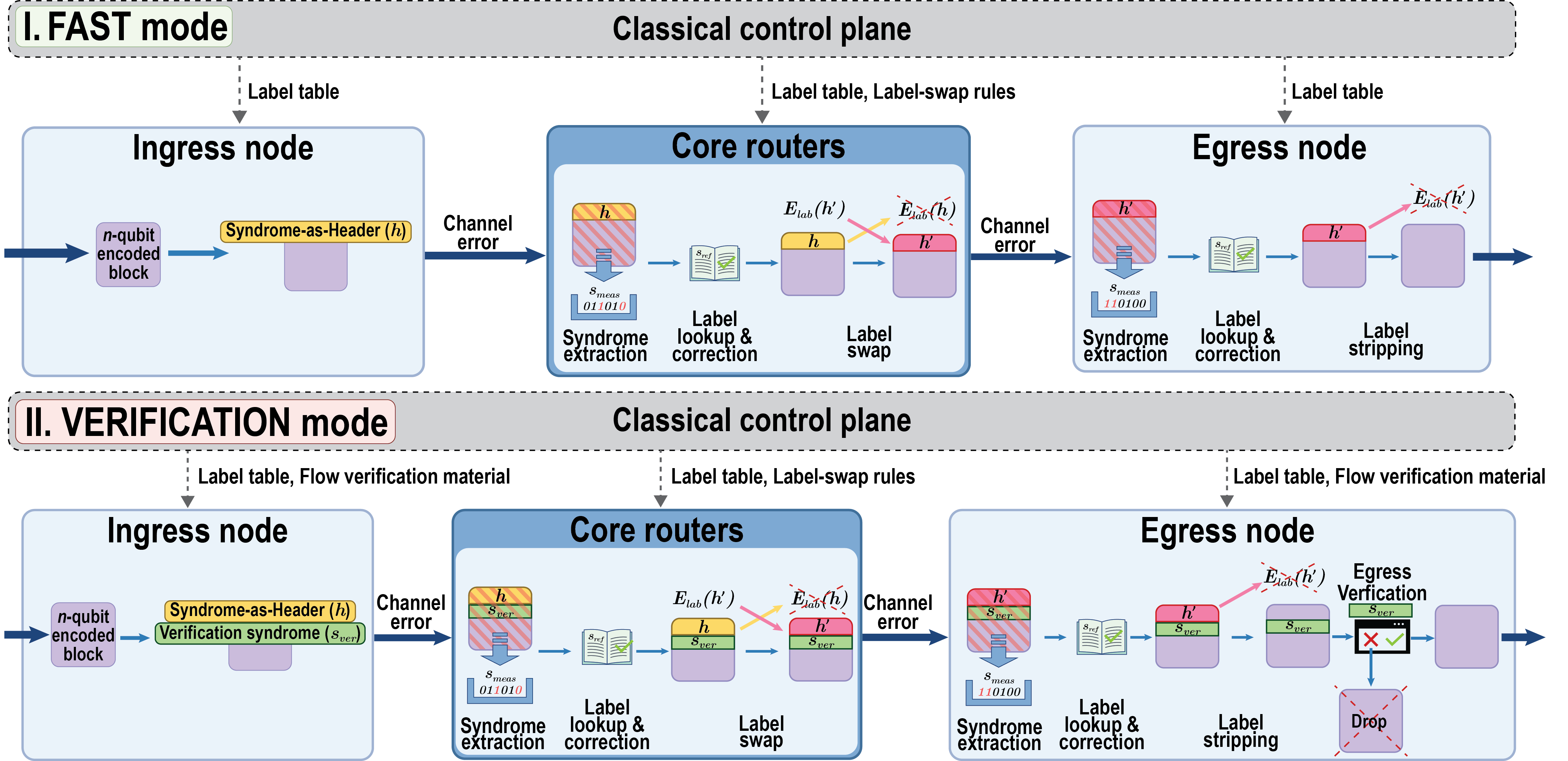}
\caption{Overview of the proposed SAH architecture in FAST and VERIFICATION modes.
The ingress embeds a per-block compact header $h$ into an encoded block via a UEI labeling operator $E_{\mathrm{lab}}(h)$ with reference syndrome $s_{\mathrm{ref}}(h)=f(h)=s(E_{\mathrm{lab}}(h))$; core routers identify the in-band header, infer the residual channel-error component, apply the corresponding correction, and swap labels without decoding the logical payload.
FAST mode performs autonomous label switching using the in-band header and pre-installed swap rules, whereas VERIFICATION mode additionally embeds a flow-specific verification component $E_{\mathrm{ver}}$ with $s_{\mathrm{ver}}=s(E_{\mathrm{ver}})$ and checks it at the egress before payload release.
The classical control plane provisions label bindings and label-swap state on a slower timescale than the quantum data plane.
Colored labels on encoded blocks denote syndrome-domain representations.}
\label{fig:sysmod}
\end{figure*}

\paragraph*{Syndrome-space header codebook}
Define the header codebook in the syndrome space as
\begin{equation}
\mathcal{C}_{\mathrm{hdr}} := \{s_{\mathrm{ref}}(h):h\in\mathcal{H}_{\mathrm{act}}\} \subseteq \mathbb{F}_2^{n-k}.
\end{equation}
For each active header $h\in\mathcal{H}_{\mathrm{act}}$, its decoding region is
\begin{equation}
s_{\mathrm{ref}}(h)\oplus\Sigma_{\mathrm{corr}} := \{s_{\mathrm{ref}}(h)\oplus s: s\in\Sigma_{\mathrm{corr}}\}.
\end{equation}
This region contains all measured syndromes that can arise from header $h$ under correctable channel-induced syndrome deviations.
A measured syndrome is assigned to header $h$ when it lies in this region.
To ensure unique header identification, the decoding regions of distinct active headers must be disjoint:
\begin{equation}
\label{eq:reg_cond}
\begin{aligned}
&\big(s_{\mathrm{ref}}(h_1)\oplus \Sigma_{\mathrm{corr}}\big)\cap \big(s_{\mathrm{ref}}(h_2)\oplus \Sigma_{\mathrm{corr}}\big)=\varnothing, \\
&\quad \forall h_1,h_2\in\mathcal{H}_{\mathrm{act}},\ h_1\neq h_2.
\end{aligned}
\end{equation}
Thus, $\mathcal{C}_{\mathrm{hdr}}$ plays the role of a code for an additive syndrome-noise channel whose error set is $\Sigma_{\mathrm{corr}}$.
Syndromes outside all decoding regions, or syndromes that are not assigned uniquely, are not valid active headers.

Because the mapping $f$ is injective over $\mathcal{H}_{\mathrm{act}}$, the number of simultaneously usable robust headers is
\begin{equation}
M_{\mathrm{hdr}}:=|\mathcal{H}_{\mathrm{act}}|.
\end{equation}
Each translated decoding region has cardinality $|\Sigma_{\mathrm{corr}}|$, and all such regions must fit disjointly inside the full syndrome space $\mathbb{F}_2^{n-k}$.
Therefore,
\begin{equation}
\label{eq:packing}
M_{\mathrm{hdr}}|\Sigma_{\mathrm{corr}}|\le 2^{n-k}.
\end{equation}
This packing bound limits the number of robust active headers and prevents all raw syndrome strings from being interpreted as usable headers.
Appendix~\ref{app:eun_ex} provides an illustrative instantiation using the $[[9,1,3]]$ Shor code and a packing-bound estimate for the $[[23,1,7]]$ Golay code.

\section{Proposed SAH protocol}
The proposed SAH protocol performs label extraction, hop-by-hop channel-error correction, label swapping, and optional end-to-end consistency verification through syndrome processing on encoded logical blocks.
Using the static state in Sec.~\ref{subsec:label_codebook}, we now describe the per-block operations in FAST and VERIFICATION modes.
Fig.~\ref{fig:sysmod} summarizes the corresponding data-plane operations and the role of the classical control plane in the two modes.

\subsection{FAST mode: autonomous label switching}
FAST mode avoids real-time per-block label signaling over the classical plane.
Routers recover the forwarding label from the in-band syndrome header and use the pre-installed forwarding table and UEI label-operator table to perform label switching autonomously.

\paragraph*{Ingress node}
For each incoming logical block $|\psi\rangle_L$, the ingress node assigns a compact header $h=(\ell,q)$, consisting of the forwarding label $\ell$ and the sequence number $q$.
It then applies the corresponding UEI labeling operator, yielding the labeled block $|\psi\rangle_{\mathrm{tx}}= E_{\mathrm{lab}}(h)|\psi\rangle_L$, which is transmitted on the quantum data plane.

\paragraph*{Core router}
Upon receiving $|\psi\rangle_{\mathrm{rec}} =E_{\mathrm{ch}}|\psi\rangle_{\mathrm{tx}}$, the router performs syndrome extraction and obtains a measured syndrome $s_{\mathrm{meas}}$.
It identifies the in-band header by checking the local header table as described in Sec.~\ref{subsec:label_codebook}.
For each candidate header $\tilde h$, the router computes
\begin{equation} \Delta s(\tilde h)=s_{\mathrm{meas}}\oplus s_{\mathrm{ref}}(\tilde h).
\end{equation}
A candidate $\tilde h$ is accepted if $\Delta s(\tilde h)\in\Sigma_{\mathrm{corr}}$.
If no such candidate exists, or if multiple candidates are found, the router declares an erasure or ambiguity and discards the block.
When a unique candidate is accepted, we denote it by $h$ and apply the corresponding public Pauli correction $C(\Delta s(h))$ to correct the channel error:
\begin{equation}
\label{eq:core_clean} |\psi\rangle_{\mathrm{clean}} = C(\Delta s(h))|\psi\rangle_{\mathrm{rec}} = E_{\mathrm{lab}}(h)|\psi\rangle_L.
\end{equation}
The router then computes the forwarding decision using only the forwarding-label subfield $\ell$, i.e., $(p,\ell')=\pi_r(\ell)$.
It preserves the sequence number $q$, constructs the outgoing compact header $h'=(\ell',q)$, and performs label swapping by applying $E_{\mathrm{swap}}=E_{\mathrm{lab}}(h')E_{\mathrm{lab}}(h)^{\dagger}$.
Since both $E_{\mathrm{lab}}(h)$ and $E_{\mathrm{lab}}(h')$ are chosen as logical-neutral representatives, this swap replaces the syndrome-domain label without introducing a nontrivial logical Pauli operation on the encoded payload, so that
\begin{equation} E_{\mathrm{swap}}|\psi\rangle_{\mathrm{clean}}=E_{\mathrm{lab}}(h')|\psi\rangle_L.
\end{equation}
The updated block is forwarded to the next hop through output port $p$.

\paragraph*{Egress node}
The egress node performs the same header identification and residual-syndrome correction as a core router, recovering $|\psi\rangle_{\mathrm{clean}} = E_{\mathrm{lab}}(h)|\psi\rangle_L$ as in~\eqref{eq:core_clean}.
It then removes the embedded label component by applying $E_{\mathrm{lab}}(h)^{\dagger}$, which restores the logical payload since $E_{\mathrm{lab}}(h)^{\dagger}E_{\mathrm{lab}}(h)|\psi\rangle_L = |\psi\rangle_L$.
The egress therefore recovers the logical payload block without requiring decoding at intermediate routers.
Using the sequence field when needed, the egress checks the block order within the flow and delivers the recovered payload to the destination-side quantum interface.

\subsection{VERIFICATION mode: end-to-end consistency verification with routable labels}
VERIFICATION mode augments FAST mode by adding a flow-specific UEI verification component that is checked at the egress before payload release, while keeping labels visible and swappable at transit routers.

\paragraph*{Control-plane state}
In VERIFICATION mode, the ingress and egress share a flow-specific UEI verification operator $E_{\mathrm{ver}}$, whose syndrome is defined as $s_{\mathrm{ver}}=s(E_{\mathrm{ver}})$.
The Pauli representative $E_{\mathrm{ver}}$ is shared only between the ingress and egress; transit routers receive only the shifted syndrome table needed for forwarding, not the representative itself.
The control plane distributes a shifted label table that maps each compact header $h$ to
\begin{equation}
s_{\mathrm{ref}}^{(\mathrm{ver})}(h)=s_{\mathrm{ref}}(h)\oplus s_{\mathrm{ver}}.
\end{equation}
The flow-specific shift is chosen so that the shifted reference syndromes remain admissible.
Pairwise disjointness of the decoding regions is preserved because the same syndrome shift is applied to all headers in the flow.
Let $A_h=s_{\mathrm{ref}}(h)\oplus\Sigma_{\mathrm{corr}}$ and $A_h^{(\mathrm{ver})}=s_{\mathrm{ref}}^{(\mathrm{ver})}(h)\oplus\Sigma_{\mathrm{corr}}$, so that $A_h^{(\mathrm{ver})}=s_{\mathrm{ver}}\oplus A_h$ and, for $h_1\neq h_2$,
\begin{equation}
A_{h_1}^{(\mathrm{ver})}\cap A_{h_2}^{(\mathrm{ver})}=s_{\mathrm{ver}}\oplus(A_{h_1}\cap A_{h_2})=\varnothing.
\end{equation}
The shifted table thus provides the syndrome information required for forwarding, while the representative $E_{\mathrm{ver}}$ remains shared only between the ingress and egress.

\paragraph*{Ingress node}
The ingress node applies both the public UEI labeling operator $E_{\mathrm{lab}}(h)$ and the ingress--egress verification operator $E_{\mathrm{ver}}$, where the latter is also chosen as a logical-neutral UEI representative, i.e., $E_{\mathrm{ver}}\in T(s_{\mathrm{ver}})\mathcal{S}$.
The verification component therefore introduces the intended syndrome shift without applying a nontrivial logical Pauli operation to the encoded payload, yielding the transmitted block $|\psi\rangle_{\mathrm{tx}} = E_{\mathrm{lab}}(h)E_{\mathrm{ver}}|\psi\rangle_L$.
Routers use the shifted table $s_{\mathrm{ref}}^{(\mathrm{ver})}(h)$ for header identification and label processing.

\paragraph*{Core router}
The core-router operation follows FAST mode with the shifted reference table.
After syndrome extraction, the router tests each candidate header $\tilde h$ against the shifted decoding region $s_{\mathrm{meas}}\in s_{\mathrm{ref}}^{(\mathrm{ver})}(\tilde h)\oplus\Sigma_{\mathrm{corr}}$.
If exactly one candidate satisfies this condition, it is denoted by $h$; otherwise, the block is discarded.
The router then computes the residual channel-error syndrome $\Delta s^{(\mathrm{ver})}(h)=s_{\mathrm{meas}}\oplus s_{\mathrm{ref}}^{(\mathrm{ver})}(h)$, applies the public correction $C(\Delta s^{(\mathrm{ver})}(h))$, and performs the same label-swap operation as in FAST mode, with $h'=(\ell',q)$ and $(p,\ell')=\pi_r(\ell)$.
After the swap, the block becomes $E_{\mathrm{lab}}(h')E_{\mathrm{ver}}|\psi\rangle_L$, so the verification component is preserved end to end while the routable label is updated hop by hop.
The updated block is forwarded to the next hop through output port $p$.

\paragraph*{Egress node}
The egress node performs the same header identification and channel-error correction as a core router.
After a unique final header $h=(\ell,q)$ is accepted and the channel error is corrected, the egress obtains $|\psi\rangle_{\mathrm{clean}}=E_{\mathrm{lab}}(h)E_{\mathrm{ver}}|\psi\rangle_L$.
The egress first checks that the recovered forwarding label $\ell$ matches the expected final-hop label for the flow.
If sequence checking is enabled, the egress also checks that $q$ lies within the expected per-flow ordering window.
Only after these header checks does the egress remove the final label component and the verification component by applying $E_{\mathrm{ver}}^{\dagger}E_{\mathrm{lab}}(h)^{\dagger}$:
\begin{equation}
|\psi\rangle_{\mathrm{strip}}=E_{\mathrm{ver}}^{\dagger}E_{\mathrm{lab}}(h)^{\dagger}E_{\mathrm{lab}}(h)E_{\mathrm{ver}}|\psi\rangle_L=|\psi\rangle_L.
\end{equation}
The egress then performs syndrome extraction and accepts the block if and only if the all-zero syndrome is obtained; otherwise, it discards the block~\cite{so25}.
A non-zero syndrome at this step indicates an uncorrected channel error or an inconsistent verification component.
Payload release occurs only if both the final-header checks and the zero-syndrome verification condition are satisfied.

\vspace{0.25em}
\noindent
\textbf{Remark:} FAST mode prioritizes autonomous forwarding under control-plane jitter by eliminating real-time per-block classical coordination during switching.
VERIFICATION mode adds end-to-end consistency verification while preserving routable labels, at the cost of additional edge-side processing and flow verification material.
The zero-syndrome verification step lets the egress detect blocks that are inconsistent with the expected ingress--egress component, including injected or misrouted blocks, though it does not constitute a cryptographic authentication guarantee.
A full information-theoretic authentication layer is outside the scope of this paper and is left to future work; the dummy-state insertion and random-permutation construction of the original UEI framework~\cite{so25} provides a natural basis for such an extension, without altering the syndrome-space label-switching mechanism.

\section{Performance evaluation}
\label{subsec:ev}
\subsection{Evaluation setup}
\begin{figure}[t]
\centering
\includegraphics[width=\linewidth]{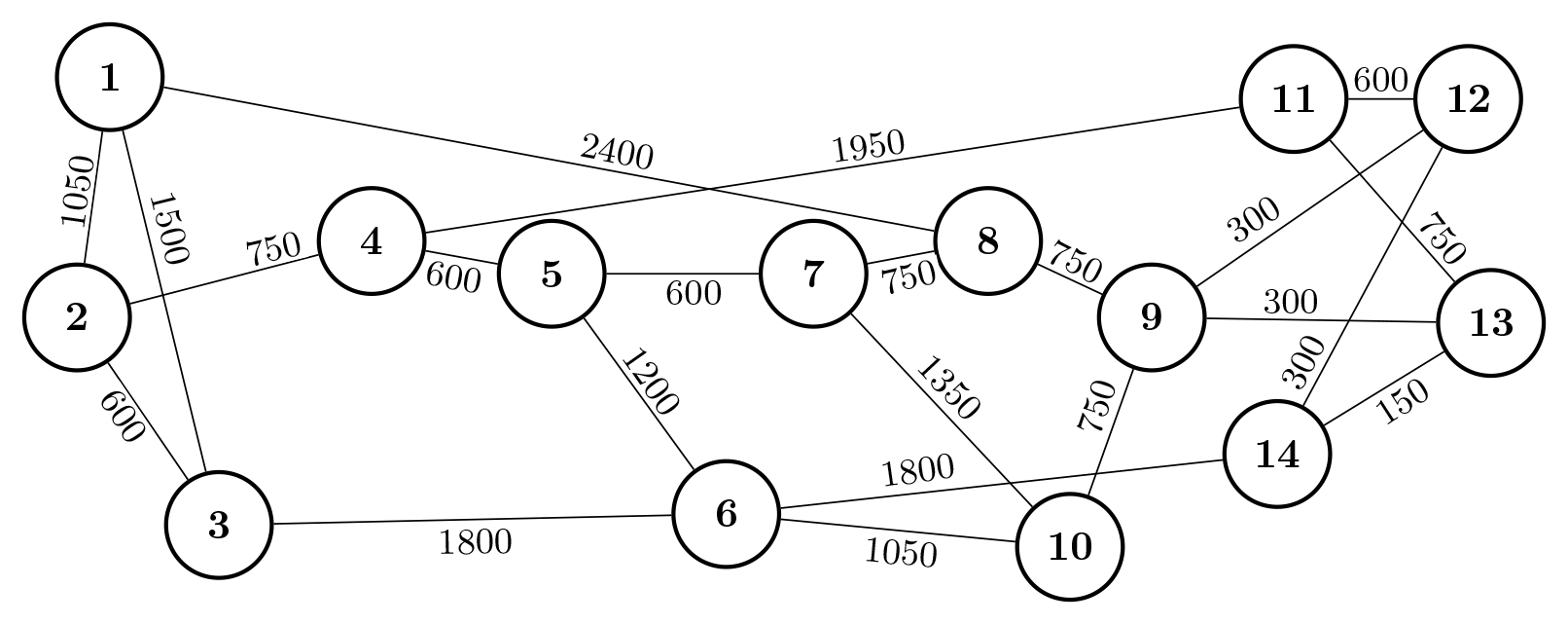}
\caption{NSFNet topology used in the performance evaluation. Edge labels indicate link lengths in kilometers.}
\label{fig:top}
\end{figure}

We evaluate the proposed SAH architecture under control-plane jitter using a QEC-normalized network-layer processing-delay benchmark on the 14-node NSFNet topology shown in Fig.~\ref{fig:top}.
For each generated payload block, the source--destination pair and shortest-path route are fixed before transmission.
The resulting traffic trace and route sequence are replayed across OBS+QEC, QW+QEC, and SAH, keeping the offered load and path distribution identical for the compared schemes.
The benchmark therefore probes how the different forwarding models respond to router-side timing uncertainty under the same network conditions.

The simulator models forwarding at the protocol layer.
Syndrome extraction, label lookup, channel-error recovery, label swapping, and verification are represented as processing stages with representative delay parameters~\cite{di22, qi99, co21, wu21}.
This abstraction avoids committing to a specific qubit modality, syndrome-extraction circuit, decoder implementation, or physical noise model, while retaining the protocol-stage delays relevant to the compared forwarding schemes.

The main benchmark isolates header--payload timing coupling in a QEC-enabled forwarding regime.
Although the system model admits per-hop Pauli errors, the main benchmark does not sample stochastic Pauli-error events, photon loss, or state-fidelity evolution for any compared scheme.
The common QEC stages, including syndrome extraction and recovery, are included as processing-delay components rather than as a sampled physical-noise-and-decoder simulation.
Because OBS+QEC, QW+QEC, and SAH share the same per-hop QEC layer, code-limited Pauli-error acceptance is treated as a common physical factor and analyzed separately in Appendix~\ref{app:qec}.
The main curves therefore focus on timing-budget loss, SAH memory residence, and delivered latency.

Propagation delay is also excluded from the main timing curves because distance-dependent propagation adds a large path-dependent baseline that can mask the microsecond-scale effect of control-plane jitter.
A propagation-inclusive evaluation using the same NSFNet topology and link lengths is provided in Appendix~\ref{app:top}.

This benchmark is performed at the payload-block level, so packet-level reassembly, partial-block loss handling, and application-dependent multi-block recovery policies are outside its scope.
For each jitter value and each switching scheme, the simulation is repeated over 30 independent random seeds.
Unless otherwise stated, the reported values are seed-averaged results, and the error bars denote the sample standard deviation across the 30 independent seeds.

\subsection{Simulation model and compared forwarding models}
\paragraph*{Network and traffic model}
The ingress generates payload blocks according to a Poisson arrival process with aggregate rate $\lambda$.
Each generated payload block is assigned a source--destination pair and a shortest-path route on the NSFNet topology.

\paragraph*{Control-plane jitter}
At each router, the router-side electronic processing time is modeled as
\begin{equation}
T_{\mathrm{cp}} = T_{\mathrm{cp}}^{\min} + |\mathcal{N}(0,\sigma_{\mathrm{cp}})|.
\end{equation}
Here, $T_{\mathrm{cp}}^{\min}$ is the minimum processing time and $\sigma_{\mathrm{cp}}$ controls the jitter level.
For OBS+QEC and QW+QEC, this term represents the jittered processing time of the external classical header used for payload alignment.
For SAH, the same sampled term represents local control-processing variability at a memory-assisted router.
A payload block may wait at a router when the router is still processing a previous block.
This waiting time is included in the delivered latency.

\paragraph*{Timing-budget model}
For OBS+QEC and QW+QEC, router-side waiting time consumes the timing budget used to align the quantum payload with the external-header control decision.
For a payload block with router waiting time $W$, the timing quantity used for baseline drop decisions is
\begin{equation}
T_{\mathrm{timing}} = T_{\mathrm{cp}} + W.
\end{equation}
In the OBS+QEC model, a guard window $G$ is used for the external control-header decision.
A payload block is accepted at a router only if
\begin{equation}
T_{\mathrm{timing}} \leq G.
\end{equation}
Otherwise, the block is discarded.

In the QW+QEC model, the payload block is held using a quantized fiber-delay-line (FDL) budget while the wrapper header is processed.
Let $\Delta_{\mathrm{FDL}}$ denote the FDL unit delay and let $K_{\mathrm{FDL}}$ denote the number of available delay classes.
The available quantized delay budgets are indexed as
\begin{equation}
d_k = k\Delta_{\mathrm{FDL}}, \qquad k=0,1,\ldots,K_{\mathrm{FDL}}-1.
\end{equation}
For a payload block with timing requirement $T_{\mathrm{timing}}$, QW+QEC selects the smallest index $k$ such that $d_k \geq T_{\mathrm{timing}}$.
If no such $k$ exists, the block is discarded.

In the SAH model, the forwarding label is already represented in the syndrome domain of the encoded block.
Accordingly, the adopted timing-budget model does not assign OBS-style guard-window drops or QW-style FDL-alignment drops to SAH.
Control-processing variability is reflected through router-side waiting, local processing time, and memory residence.

\paragraph*{QEC-normalized processing}
The QEC-normalized baselines include the same per-hop QEC processing component
\begin{equation}
T_{\mathrm{QEC}} = T_{\mathrm{syn}} + T_{\mathrm{rec}}.
\end{equation}
For OBS+QEC, QEC processing on the arriving encoded payload block is modeled as a serial per-hop latency cost.
For an accepted payload block $i$ traversing $H_i$ hops, the delivered latency is
\begin{equation}
L_i^{\mathrm{OBS+QEC}} = L_i^{\mathrm{OBS\ timing}} + H_iT_{\mathrm{QEC}}.
\end{equation}
For QW+QEC, QEC processing is assumed to overlap with the wrapper/header-processing interval and is not added as a separate serial latency term:
\begin{equation}
L_i^{\mathrm{QW+QEC}} = L_i^{\mathrm{QW\ timing}}.
\end{equation}
Here, $L_i^{\mathrm{OBS\ timing}}$ and $L_i^{\mathrm{QW\ timing}}$ denote the accepted-block timing latency accumulated by the OBS and QW timing-budget models before scheme-specific QEC-latency accounting.

\begin{table}[t]
\centering
\caption{Evaluation parameters and their values.}
\label{tab}
\renewcommand{\arraystretch}{1.3}
\begin{tabular}{@{}ll p{4.3cm}@{}}
\toprule
Parameter & Value & Description \\
\midrule
$\lambda$ & $4.0\times 10^{3}$ blocks/s & Aggregate block generation rate \\
$T_{\mathrm{cp}}^{\min}$ & $10~\mu\mathrm{s}$ & Min. control-plane processing time \\
$\sigma_{\mathrm{cp}}$ & $0$--$10~\mu\mathrm{s}$ & Control-plane jitter level \\
$G$ & $20~\mu\mathrm{s}$ & OBS+QEC guard window \\
$\Delta_{\mathrm{FDL}}$ & $5~\mu\mathrm{s}$ & QW+QEC fiber-delay-line unit \\
$K_{\mathrm{FDL}}$ & $5$ & Number of QW+QEC delay classes \\
$T_{\mathrm{syn}}$ & $1~\mu\mathrm{s}$ & Syndrome extraction time \\
$T_{\mathrm{rec}}$ & $1~\mu\mathrm{s}$ & Channel-error recovery time \\
$T_{\mathrm{QEC}}$ & $2~\mu\mathrm{s}$ & Per-hop QEC processing, $T_{\mathrm{syn}}+T_{\mathrm{rec}}$ \\
$T_{\mathrm{lookup}}$ & $1~\mu\mathrm{s}$ & Label lookup time \\
$T_{\mathrm{swap}}$ & $1~\mu\mathrm{s}$ & Label-swap time \\
$T_{\mathrm{ctrl}}$ & $1~\mu\mathrm{s}$ & Local control overhead \\
$T_{\mathrm{ver}}$ & $5~\mu\mathrm{s}$ & VERIFICATION-mode overhead \\
$N_{\mathrm{seed}}$ & $30$ & Random seeds per jitter--scheme point \\
\bottomrule
\end{tabular}
\end{table}

In FAST mode, the per-hop SAH processing time is modeled as
\begin{equation}
T_{\mathrm{SAH}}^{(\mathrm{F})} = T_{\mathrm{syn}} + T_{\mathrm{lookup}} + T_{\mathrm{rec}} + T_{\mathrm{swap}} + T_{\mathrm{ctrl}}.
\end{equation}
Here, $T_{\mathrm{syn}}$ and $T_{\mathrm{rec}}$ correspond to the common syndrome extraction and recovery stages, while $T_{\mathrm{lookup}}$, $T_{\mathrm{swap}}$, and $T_{\mathrm{ctrl}}$ represent SAH-specific syndrome-domain forwarding overhead.
In VERIFICATION mode, an additional verification-related processing cost is included:
\begin{equation}
T_{\mathrm{SAH}}^{(\mathrm{V})} = T_{\mathrm{SAH}}^{(\mathrm{F})} + T_{\mathrm{ver}}.
\end{equation}

\paragraph*{Memory-residence penalty}
When the memory-residence penalty is enabled for SAH, the dwell-time survival factor is modeled using the physical memory coherence time $T_2$ of router memories~\cite{bh20,az23}.
Let $W_{\mathrm{SAH}}$ denote the SAH memory-residence time associated with router-side waiting and local syndrome-domain processing.
The block survival factor is represented as
\begin{equation}
P_{\mathrm{mem}}(W_{\mathrm{SAH}})=e^{-W_{\mathrm{SAH}}/T_2}.
\end{equation}
This factor is used as a coarse $T_2$-based survival proxy for the local memory residence incurred by SAH forwarding.
The evaluation sweeps $T_2$ over a range of memory-coherence settings to show how the SAH advantage depends on router memory coherence.

\paragraph*{Evaluation parameters}
Table~\ref{tab} summarizes the parameters used in the processing-delay benchmark.
The numerical values are representative benchmark parameters chosen to expose the relative timing behavior of OBS+QEC, QW+QEC, and SAH under microsecond-scale control-plane jitter, rather than hardware-specific measurements.
Unless otherwise stated, the evaluation varies the control-plane jitter level $\sigma_{\mathrm{cp}}$ while keeping the remaining parameters fixed.
The guard-window ($G=20~\mu\mathrm{s}$) and FDL ($\Delta_{\mathrm{FDL}}=5~\mu\mathrm{s}$, $K_{\mathrm{FDL}}=5$) budgets and the SAH memory coherence ($T_2=1~\mathrm{ms}$) represent distinct physical resources.
The guard-window and FDL budgets represent microsecond-scale latency-alignment and physical-delay-line resources for external-header forwarding, whereas $T_2$ represents local memory coherence in memory-assisted forwarding nodes~\cite{bh20, az23}.

\subsection{Performance metrics}
\paragraph*{Acceptance probability}
Let $N_{\mathrm{acc}}(T)$ denote the number of payload blocks accepted at the egress during an observation window of duration $T$.
Let $N_{\mathrm{drop}}(T)$ denote the number of payload blocks discarded by timing-budget failure, syndrome erasure, ambiguity, verification failure, memory-residence rejection, or other experiment-specific rejection events.
We define the acceptance probability as
\begin{equation}
P_{\mathrm{acc}}(T) := \frac{N_{\mathrm{acc}}(T)}{N_{\mathrm{acc}}(T)+N_{\mathrm{drop}}(T)}.
\end{equation}
This metric measures the fraction of accounted payload blocks that are successfully delivered.

\paragraph*{Effective throughput}
We define the effective throughput as the accepted payload-block rate:
\begin{equation}
\Theta(T) := \frac{N_{\mathrm{acc}}(T)}{T}.
\end{equation}
Thus, the throughput metric reports the delivered rate of accepted payload blocks, not the offered generation rate.
When needed, this block-level throughput can be converted into an equivalent logical-qubit rate by multiplying by the number of logical qubits represented by each payload block.

\begin{figure*}[t]
\centering
\includegraphics[width=\linewidth]{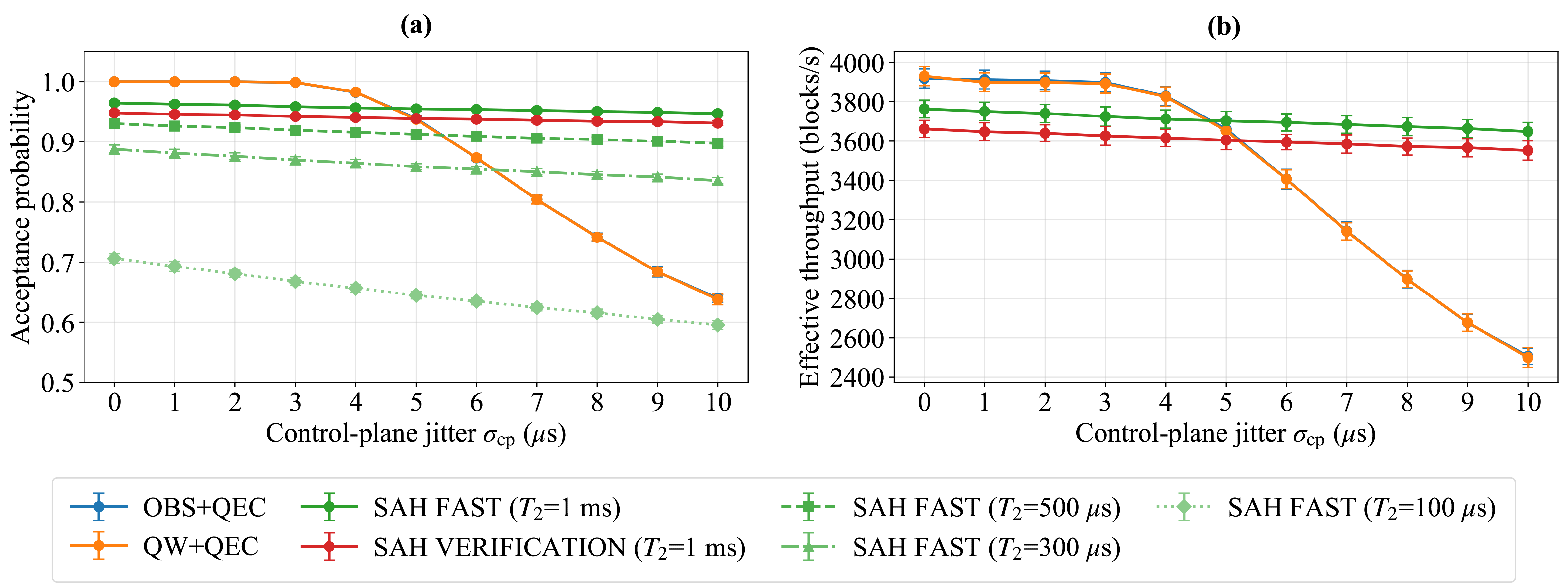}
\caption{Evaluation under control-plane jitter in the QEC-normalized timing benchmark, averaged over 30 independent random seeds.
(a) Acceptance probability.
SAH FAST is evaluated with the physical-$T_2$-mapped memory-residence penalty for $T_2\in\{1~\mathrm{ms},500~\mu\mathrm{s},300~\mu\mathrm{s},100~\mu\mathrm{s}\}$, while SAH VERIFICATION is shown at $T_2=1~\mathrm{ms}$.
(b) Effective accepted-block throughput.
The SAH curves in (b) use the $T_2=1~\mathrm{ms}$ setting.
Because all schemes use the same offered traffic trace, the throughput trend in (b) follows the acceptance behavior in (a).
OBS+QEC and QW+QEC nearly coincide in both panels.}
\label{fig:combined_jitter}
\end{figure*}

\paragraph*{Delivered latency}
For an accepted payload block $i$, let $t_i^{(\mathrm{in})}$ be the time at which the block becomes ready at the ingress.
Let $t_i^{(\mathrm{out})}$ be the logical time at which the block is accepted at the egress after router-side processing.
The delivered latency is defined as
\begin{equation}
L_i := t_i^{(\mathrm{out})}-t_i^{(\mathrm{in})}.
\end{equation}
We report the seed-averaged mean of $L_i$ over accepted payload blocks.
This metric is propagation-normalized: it includes router-side waiting, common per-hop QEC processing, and scheme-specific per-hop processing time, while excluding distance-dependent propagation through fiber links.

\subsection{Results}
\label{subsec:res}
\subsubsection{Acceptance probability and effective throughput under control-plane jitter}
Fig.~\ref{fig:combined_jitter} shows the acceptance probability and effective accepted-block throughput as functions of the control-plane jitter level.
Fig.~\ref{fig:combined_jitter}a reports the acceptance probability with the memory-residence penalty applied to SAH.
For SAH FAST, the memory coherence time is swept over $T_2 \in \{1~\mathrm{ms}, 500~\mu\mathrm{s}, 300~\mu\mathrm{s}, 100~\mu\mathrm{s}\}$.
SAH VERIFICATION is shown at the setting $T_2=1~\mathrm{ms}$.
Fig.~\ref{fig:combined_jitter}b reports the accepted-block throughput under the same offered traffic trace, using $T_2=1~\mathrm{ms}$ for the SAH curves.
Because the same traffic trace is reused across schemes, the throughput curve mainly follows the accepted-block fraction.

OBS+QEC and QW+QEC maintain acceptance probability close to one at low jitter, but their acceptance probabilities decrease as $\sigma_{\mathrm{cp}}$ increases.
At $\sigma_{\mathrm{cp}}=10~\mu\mathrm{s}$, OBS+QEC and QW+QEC drop to $0.640$ and $0.638$, respectively.
These correspond to absolute decreases of $36.0$ and $36.2$ percentage points from $\sigma_{\mathrm{cp}}=0$.
This degradation follows the increasing incidence of external-header alignment failure at higher jitter.

In contrast, the SAH curves are mainly limited by the memory-residence penalty.
For SAH FAST with $T_2=1~\mathrm{ms}$, the acceptance probability decreases from $0.965$ at $\sigma_{\mathrm{cp}}=0$ to $0.947$ at $\sigma_{\mathrm{cp}}=10~\mu\mathrm{s}$.
For SAH VERIFICATION with $T_2=1~\mathrm{ms}$, the acceptance probability decreases from $0.948$ to $0.931$ over the same jitter range.
For $T_2=1~\mathrm{ms}$, SAH FAST overtakes OBS+QEC and QW+QEC around $\sigma_{\mathrm{cp}}\approx5~\mu\mathrm{s}$.
Shorter $T_2$ values reduce the SAH FAST acceptance probability further, as shown by the $500~\mu\mathrm{s}$, $300~\mu\mathrm{s}$, and $100~\mu\mathrm{s}$ curves.
The $100~\mu\mathrm{s}$ case represents a stress condition in which the memory-residence penalty becomes comparable to the timing-budget loss of the baselines.

The same trend appears in the effective-throughput curve.
At $\sigma_{\mathrm{cp}}=0$, OBS+QEC and QW+QEC achieve $3.918\times10^{3}$ and $3.930\times10^{3}$ blocks/s, respectively.
At $\sigma_{\mathrm{cp}}=10~\mu\mathrm{s}$, their throughput decreases to $2.506\times10^{3}$ and $2.499\times10^{3}$ blocks/s, respectively.
These correspond to relative throughput decreases of $36.1\%$ and $36.4\%$.
For the $T_2=1~\mathrm{ms}$ setting, SAH FAST decreases from $3.763\times10^{3}$ to $3.649\times10^{3}$ blocks/s as $\sigma_{\mathrm{cp}}$ increases from $0$ to $10~\mu\mathrm{s}$.
SAH VERIFICATION decreases from $3.662\times10^{3}$ to $3.552\times10^{3}$ blocks/s over the same range.
These correspond to relative throughput decreases of $3.0\%$ for both SAH modes.
Thus, under the $T_2=1~\mathrm{ms}$ setting, SAH maintains a substantially flatter delivered-rate curve than OBS+QEC and QW+QEC as control-plane jitter increases.

\subsubsection{Delivered-latency cost under control-plane jitter}
\begin{figure}[t]
\centering
\includegraphics[width=\linewidth]{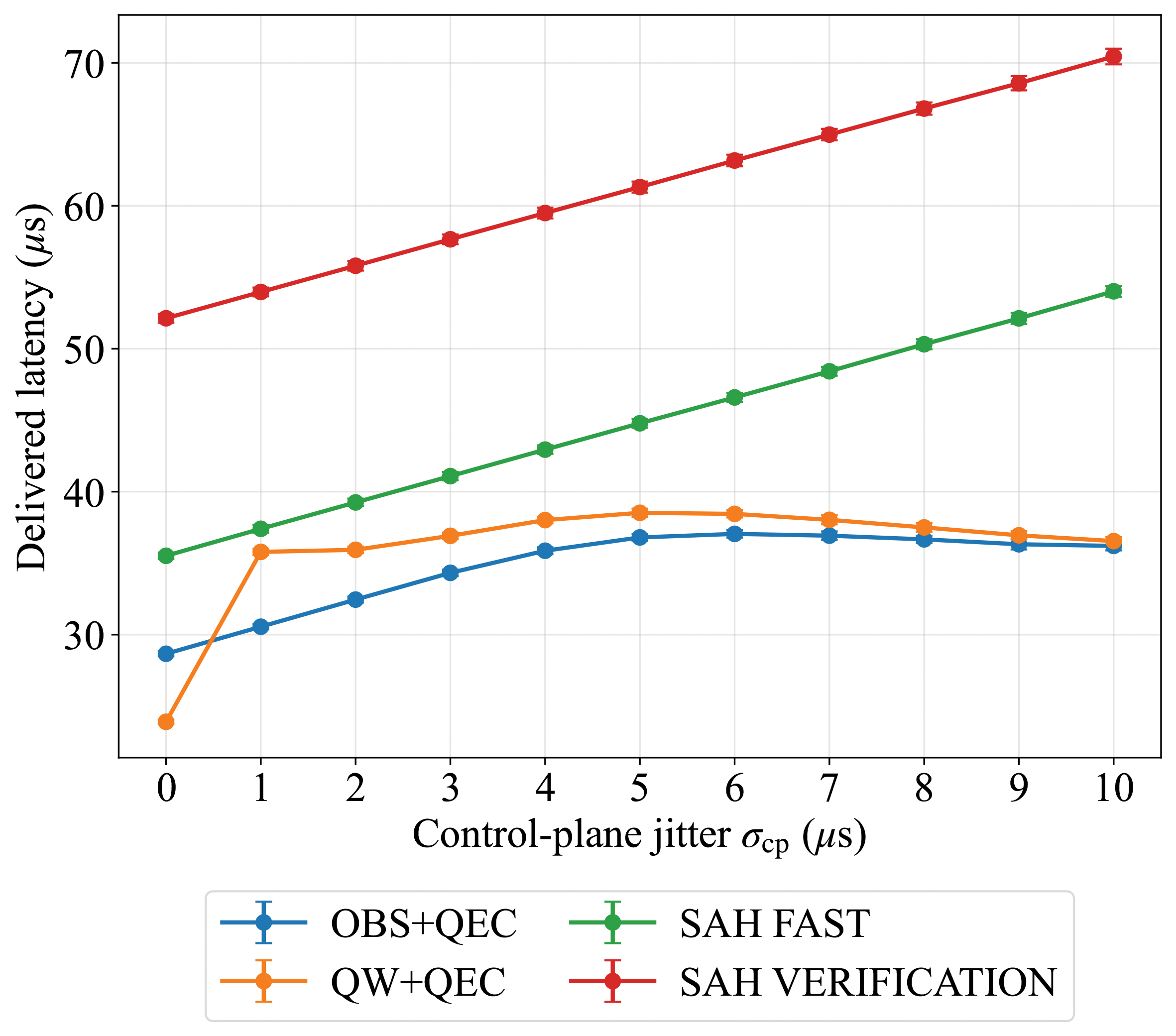}
\caption{Mean delivered latency under control-plane jitter in the QEC-normalized timing benchmark, averaged over 30 independent random seeds.
For SAH, the delivered-block population is evaluated under the $T_2=1~\mathrm{ms}$ setting.
The reported latency includes the processing stages specified in the simulation model and is averaged over accepted payload blocks.}
\label{fig:lj}
\end{figure}

Fig.~\ref{fig:lj} plots the mean delivered latency as a function of the control-plane jitter level.
The latency is averaged over delivered blocks only.
At $\sigma_{\mathrm{cp}}=0$, QW+QEC has the lowest delivered latency, $23.902~\mu\mathrm{s}$.
OBS+QEC starts at $28.660~\mu\mathrm{s}$.
SAH FAST starts at $35.515~\mu\mathrm{s}$, and SAH VERIFICATION starts at $52.130~\mu\mathrm{s}$.
These offsets reflect the processing-stage accounting defined in the simulation model.

QW+QEC exhibits an abrupt latency increase between $\sigma_{\mathrm{cp}}=0$ and $\sigma_{\mathrm{cp}}=1~\mu\mathrm{s}$, rising from $23.902~\mu\mathrm{s}$ to $35.789~\mu\mathrm{s}$.
This jump is consistent with quantized FDL allocation, where the required waiting time is rounded up to an available delay budget.
After this jump, QW+QEC increases to $38.521~\mu\mathrm{s}$ at $\sigma_{\mathrm{cp}}=5~\mu\mathrm{s}$, and then decreases to $36.542~\mu\mathrm{s}$ at $\sigma_{\mathrm{cp}}=10~\mu\mathrm{s}$.
OBS+QEC increases to $37.046~\mu\mathrm{s}$ at $\sigma_{\mathrm{cp}}=6~\mu\mathrm{s}$, and then decreases to $36.208~\mu\mathrm{s}$ at $\sigma_{\mathrm{cp}}=10~\mu\mathrm{s}$.
These high-jitter decreases do not indicate improved timing behavior.
They occur because timing-constrained blocks are increasingly discarded, and the latency average is computed only over the accepted-block population.

For SAH, the same delivered-block conditioning applies.
However, at $T_2=1~\mathrm{ms}$, the memory-drop fraction is small, so the reported SAH latency is only weakly affected by this survivor effect.
SAH FAST increases from $35.515~\mu\mathrm{s}$ at $\sigma_{\mathrm{cp}}=0$ to $54.016~\mu\mathrm{s}$ at $\sigma_{\mathrm{cp}}=10~\mu\mathrm{s}$.
SAH VERIFICATION increases from $52.130~\mu\mathrm{s}$ to $70.439~\mu\mathrm{s}$ over the same range.
The SAH curves increase with $\sigma_{\mathrm{cp}}$ because accepted blocks retain router-side waiting, local processing, and memory-residence delay.

Overall, the QEC-normalized jitter results show a timing--memory--latency trade-off.
Under the $T_2=1~\mathrm{ms}$ setting, SAH maintains flatter delivered-rate curves than OBS+QEC and QW+QEC, at the cost of higher delivered latency and dependence on the physical-$T_2$-mapped memory-residence penalty.

\section{Conclusion}
This paper introduced SAH, a quantum label-switching architecture that embeds forwarding information into the syndrome structure of an encoded quantum payload.
Using UEI-induced reference syndromes and a syndrome-space header codebook, the proposed scheme enables routers to identify labels, suppress correctable channel-error deviations, and swap labels, operating purely in the syndrome domain rather than on the logical payload or an external classical header.
Under the QEC-normalized timing benchmark, OBS+QEC and QW+QEC remain limited by guard-window or fiber-delay-line timing-budget failures under control-plane jitter, whereas SAH reflects the same timing uncertainty as delivered latency and memory residence.

By construction, SAH removes the need to align an external header with an in-flight quantum payload, thereby eliminating external-header timing-budget drops under the adopted model.
At $\sigma_{\mathrm{cp}}=10~\mu\mathrm{s}$ and $T_2=1~\mathrm{ms}$, SAH FAST achieves an acceptance probability of $0.947$ and a throughput of $3.649\times10^3$ blocks/s, compared with about $0.64$ and $2.5\times10^3$ blocks/s for OBS+QEC and QW+QEC.
Its advantage depends on router memory coherence: with sufficiently long coherence, SAH maintains flatter acceptance and throughput curves, whereas the advantage is reduced when memory-residence loss becomes dominant.
Thus, the proposed scheme converts external-header jitter-induced delivery loss into memory-dependent residence loss and increased delivered latency from local syndrome-domain processing.

The physical-error sensitivity analysis in Appendix~\ref{app:qec} further shows that channel-error-induced losses are governed by the selected QECC, the physical error model, and the route hop count.
Accordingly, the timing advantage of SAH should be read as the removal of an alignment bottleneck, not as immunity to physical channel errors.
Future work will calibrate the present timing-and-code-limited framework to hardware-specific channel, memory, and syndrome-processing models, and extend the block-level benchmark to packet-level recovery.

\appendices
\section{Illustrative syndrome-space codebook construction and header-budget estimate}
\label{app:eun_ex}
\subsection{Shor-code convention and stabilizer generators}
This appendix gives an explicit syndrome-space codebook example for the Shor $[[9,1,3]]$ QECC.
The example illustrates the syndrome-labeling construction and the resulting header-budget constraint, rather than optimizing the codebook size.
We fix the qubit index order from left to right as $1,2,\dots,9$ in each Pauli string.
The syndrome is written as $s(E)=(s_X(E)\,|\,s_Z(E))\in\{0,1\}^8$, where the first two bits correspond to the $X$-type stabilizer measurements and the last six bits correspond to the $Z$-type stabilizer measurements under the generator ordering below.
All listed reference syndromes and label operators are defined relative to this fixed convention.
We use the following independent stabilizer generators, with the first two being $X$-type and the last six being $Z$-type:
\begin{equation}
\begin{aligned}
g_1^{(X)} &= XXXXXXIII, & g_2^{(X)} &= IIIXXXXXX, \\
g_1^{(Z)} &= ZZIIIIIII, & g_2^{(Z)} &= IZZIIIIII, \\
g_3^{(Z)} &= IIIZZIIII, & g_4^{(Z)} &= IIIIZZIII, \\
g_5^{(Z)} &= IIIIIIZZI, & g_6^{(Z)} &= IIIIIIIZZ.
\end{aligned}
\end{equation}
The syndrome bits are ordered as
\begin{equation}
\begin{aligned}
s(E)=(&s(g_1^{(X)};E),s(g_2^{(X)};E),\,|\\
&s(g_1^{(Z)};E),s(g_2^{(Z)};E),s(g_3^{(Z)};E),\\
&s(g_4^{(Z)};E),s(g_5^{(Z)};E),s(g_6^{(Z)};E)).
\end{aligned}
\end{equation}
Here, $s(g;E)=1$ iff $E$ anticommutes with $g$.
The listed reference syndromes are computed from this stabilizer ordering using the standard symplectic commutation rule.

\subsection{Illustrative syndrome-space header codebook}
For this Shor-code convention, the correctable single-qubit syndrome set is
\begin{equation}
\Sigma_{\mathrm{corr}} = \{s(I)\}\cup\{s(X_i),s(Z_i),s(Y_i): i=1,\ldots,9\}.
\end{equation}
The following four-entry syndrome-space header codebook gives a compact illustrative example:
\begin{equation}
\begin{array}{c|c|c}
h & s_{\mathrm{ref}}(h) & E_{\mathrm{lab}}(h) \\
\hline
00 & 11\,|\,000001 & Z_6X_7X_8 \\
01 & 11\,|\,010100 & Z_4X_1X_2X_6 \\
10 & 11\,|\,101010 & X_2X_3Y_4X_7 \\
11 & 11\,|\,111111 & Z_4X_1X_3X_5X_8
\end{array}
\end{equation}
Each listed representative realizes the reference syndrome assigned to the corresponding header in the public lookup table.
The representatives are not unique; however, only representatives in the logical-neutral class $T(s_{\mathrm{ref}}(h))\mathcal{S}$ are admissible for SAH labeling.
Representatives with the same reference syndrome but differing by multiplication with a nontrivial logical Pauli operator are excluded, because they would transform the encoded logical payload.
The four reference syndromes are selected outside the correctable single-qubit syndrome region and satisfy the disjoint decoding-region condition defined in Sec.~\ref{subsec:label_codebook}.
Thus, correctable channel-induced syndrome deviations around one reference syndrome are not confused with the decoding region of another reference syndrome.

\subsection{Header-budget estimate for a larger code}
We apply the packing bound~\eqref{eq:packing} to the $[[23,1,7]]$ Golay code to estimate the number of robust active headers supported by a larger syndrome space.
For an $[[n,k,d]]$ QECC with guaranteed correction radius $t=\lfloor(d-1)/2\rfloor$, the number of Pauli patterns of weight at most $t$ is
\begin{equation}
B_t(n):=\sum_{w=0}^{t}3^{w}\binom{n}{w}.
\end{equation}
For the nondegenerate Golay code considered here, distinct Pauli errors of weight at most $t$ have distinct syndromes, so this count coincides with $\lvert\Sigma_{\mathrm{corr}}\rvert$.
For a degenerate code, the actual value of $\lvert\Sigma_{\mathrm{corr}}\rvert$ should be used in the packing bound instead, while $B_t(n)$ gives a conservative replacement.
Thus, the number of simultaneously used active headers must satisfy
\begin{equation}
\label{eq:bpack}
M_{\mathrm{hdr}}\le M_{\mathrm{hdr}}^{\mathrm{pack}},\qquad M_{\mathrm{hdr}}^{\mathrm{pack}} := \left\lfloor \frac{2^{n-k}}{B_t(n)} \right\rfloor.
\end{equation}

\begin{table}[t]
\centering
\caption{Packing-level robust-header-count estimate for the $[[23,1,7]]$ Golay code.}
\label{tab:budget}
\renewcommand{\arraystretch}{1.4}
\begin{tabular}{lcccc}
\hline
Code & $n-k$ & $t$ & $B_t(n)$ & $M_{\mathrm{hdr}}^{\mathrm{pack}}$ \\
\hline
Golay $[[23,1,7]]$ & $22$ & $3$ & $50164$ & $83$ \\
\hline
\end{tabular}
\end{table}
Table~\ref{tab:budget} shows that the raw syndrome dimension alone can substantially overstate the number of robustly usable active headers.
Although the $[[23,1,7]]$ Golay code has $2^{22}$ possible syndrome strings, the packing-level bound allows at most $83$ active reference syndromes with disjoint correctable-syndrome neighborhoods.
Thus, this code is sufficient when the switching core requires no more than $83$ active compact headers.
Larger active header spaces require codes with more syndrome headroom relative to $B_t(n)$, or an extended multi-block header-framing design.

\section{Simulation with propagation delay}
\label{app:top}
This appendix reports a propagation-inclusive latency check using the same NSFNet topology, traffic traces, routes, and router-side timing parameters as in Sec.~\ref{subsec:ev}.
Unlike the main QEC-normalized timing benchmark, this check includes distance-dependent propagation delay in the delivered-latency calculation.
For a link of length $d$, the propagation delay is computed as $\tau=d/v$, where $v=2\times10^{5}~\mathrm{km/s}$ is the propagation speed in optical fiber.
Stochastic Pauli-error events, photon loss, and state-fidelity tracking are excluded, as in the main timing benchmark.
The acceptance-probability and effective-throughput curves are not repeated because propagation delay changes the delivered-latency scale but does not change the local timing-budget drop rules used by OBS+QEC and QW+QEC, nor does it introduce an external-header timing-budget drop for SAH.

\begin{figure}[h]
\centering
\includegraphics[width=\columnwidth]{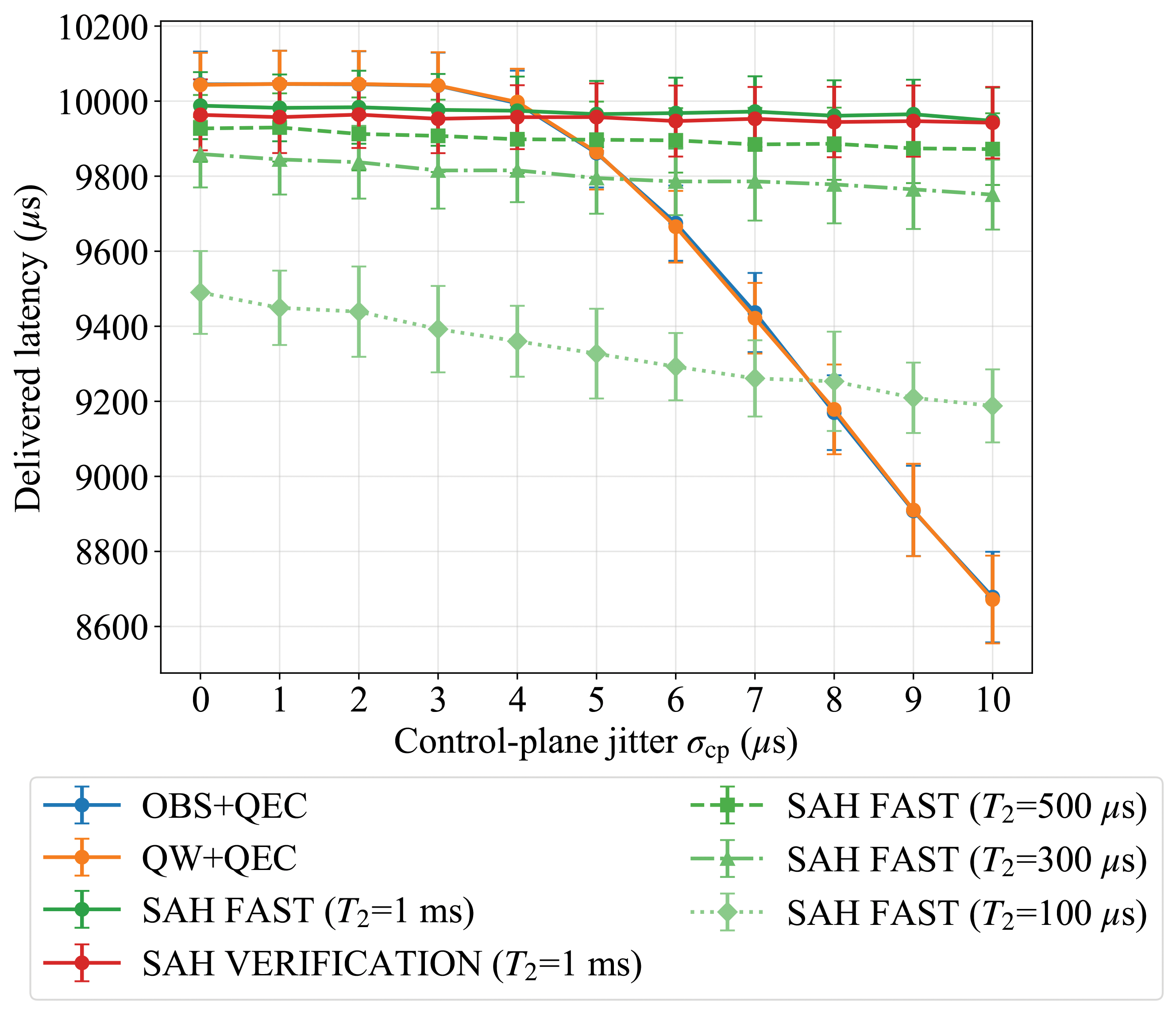}
\caption{Mean delivered latency under control-plane jitter in the propagation-inclusive QEC-normalized benchmark.
Distance-dependent propagation shifts the latency scale from the tens-of-microseconds range in Fig.~\ref{fig:lj} to the millisecond-scale range.
The high-jitter decrease of OBS+QEC and QW+QEC is a survivor-bias effect.}
\label{fig:ljpd}
\end{figure}

Fig.~\ref{fig:ljpd} shows that propagation delay dominates the absolute delivered-latency scale.
As a result, the relative contribution of SAH's syndrome-domain processing overhead becomes smaller in the propagation-inclusive setting.
At high jitter, OBS+QEC and QW+QEC preferentially discard timing-constrained blocks, so the latency average is computed only over a biased accepted-block population.
Thus, the propagation-inclusive result supports the main QEC-normalized timing benchmark: propagation changes the absolute latency scale, while the robustness difference remains governed by the presence or absence of external-header timing-budget-induced drops.

\section{Code-limited acceptance under Pauli errors}
\label{app:qec}
\begin{figure}[t]
\centering
\includegraphics[width=\linewidth]{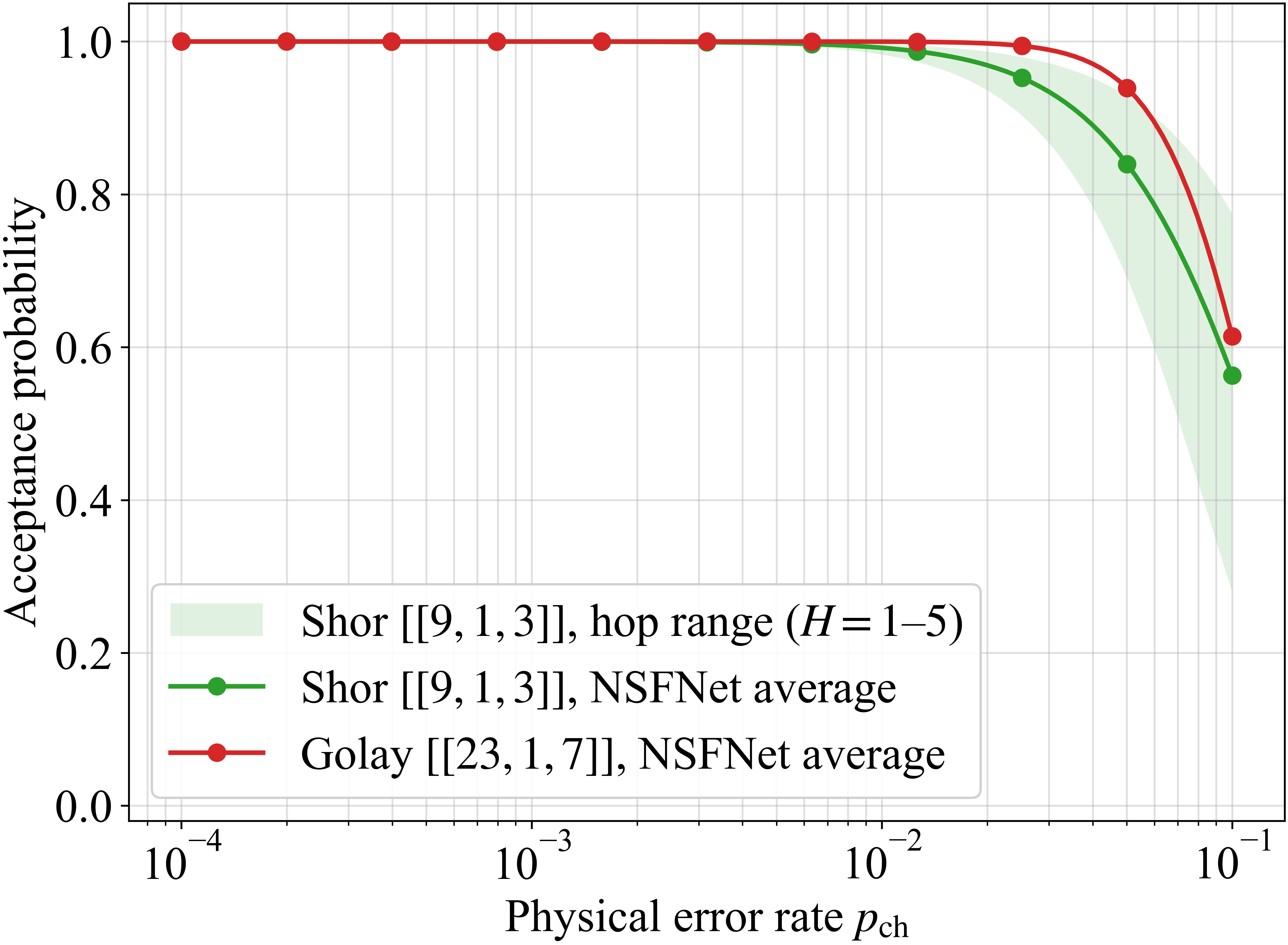}
\caption{Code-limited acceptance probability under physical errors in an independent depolarizing channel.
Solid curves show the NSFNet route-averaged acceptance estimate for Shor $[[9,1,3]]$ and Golay $[[23,1,7]]$ QECCs, and the shaded region shows the full shortest-path hop-count range for the Shor-code case.
The estimate depends on the selected code and route hop count under QEC-normalized encoded-block forwarding.}
\label{fig:qec_acc}
\end{figure}

Fig.~\ref{fig:qec_acc} complements the QEC-normalized timing benchmark by estimating the code-limited acceptance probability under physical channel errors.
Whereas the preceding jitter-based results isolate timing-induced drops caused by external-header alignment constraints, Fig.~\ref{fig:qec_acc} isolates the acceptance loss caused by Pauli errors that exceed the correction capability of the selected QECC.
Because OBS+QEC, QW+QEC, and SAH use the same per-hop QEC layer in the main benchmark, this channel-error effect is treated as a common code-limited factor; hence the near-unit SAH acceptance in the jitter curves should be read as the removal of guard-window and FDL timing-budget failures, not as immunity to channel-error-induced losses.
We assume an independent depolarizing channel in which each physical qubit undergoes a non-identity Pauli error with probability $p_{\mathrm{ch}}$ per hop.
For an $[[n,k,d]]$ QECC with guaranteed correction radius $t=\lfloor(d-1)/2\rfloor$, let $P_{\le t}(p_{\mathrm{ch}})$ denote the probability that a single encoded block incurs at most $t$ physical errors,
\begin{equation}
\label{eq:pcw}
P_{\le t}(p_{\mathrm{ch}}) = \sum_{w=0}^{t} \binom{n}{w} p_{\mathrm{ch}}^{w} (1-p_{\mathrm{ch}})^{n-w}.
\end{equation}

This correctable-weight probability yields a conservative acceptance estimate: it treats every error of weight at most $t$ as correctable and every higher-weight error as a failure, thereby neglecting the residual correction of some higher-weight errors by degenerate codes or particular decoder choices.
Assuming independent per-hop physical errors and per-hop correction, a route with $H$ hops has conservative route-level acceptance estimate $P_{\le t}(p_{\mathrm{ch}})^{H}$, and the plotted solid curves report the NSFNet route-average
\begin{equation}
\label{eq:asah}
A_{\mathrm{QEC}}(p_{\mathrm{ch}}) = \frac{1}{N}\sum_{i=1}^{N} P_{\le t}(p_{\mathrm{ch}})^{H_i} = \sum_{H} \pi_H P_{\le t}(p_{\mathrm{ch}})^{H},
\end{equation}
over all $N$ ordered source--destination pairs, where $H_i$ is the hop count of pair $i$ and $\pi_H$ is the shortest-path hop-count distribution.
The Shor-code curve provides a small-code reference with guaranteed correction radius $t=1$, while the Golay-code curve provides a stronger-code reference with guaranteed correction radius $t=3$.
As expected, both curves remain close to unity at low physical error rates, and the acceptance decreases as $p_{\mathrm{ch}}$ increases.
The decrease is faster for Shor $[[9,1,3]]$, whereas Golay $[[23,1,7]]$ maintains higher acceptance over the same error-rate range.
The shaded region for the Shor-code case shows the effect of route length across the NSFNet shortest-path distribution.
Accordingly, Fig.~\ref{fig:qec_acc} should be read as the common physical-error acceptance ceiling imposed by the selected QECC and route hop count, complementing the timing-induced acceptance and throughput behavior isolated in the preceding jitter-based figures.


\begin{thebibliography}{10}
\bibitem{fo13}
B.~A. Forouzan, {\em Data Communications and Networking}.
\newblock McGraw-Hill, 5~ed., 2013.
\newblock Global Edition.

\bibitem{ki08}
H.~J. Kimble, ``The quantum internet,'' {\em Nature}, vol.~453, no.~7198, pp.~1023--1030, 2008.

\bibitem{we18}
S.~Wehner, D.~Elkouss, and R.~Hanson, ``Quantum internet: A vision for the road ahead,'' {\em Science}, vol.~362, no.~6412, p.~eaam9288, 2018.

\bibitem{ca19}
A.~S. Cacciapuoti, M.~Caleffi, F.~Tafuri, F.~S. Cataliotti, S.~Gherardini, and G.~Bianchi, ``Quantum internet: Networking challenges in distributed quantum computing,'' {\em IEEE Network}, vol.~34, no.~1, pp.~137--143, 2019.

\bibitem{rv14}
R.~Van~Meter, {\em Quantum networking}.
\newblock John Wiley \& Sons, 2014.

\bibitem{da19}
A.~Dahlberg, M.~Skrzypczyk, T.~Coopmans, L.~Wubben, F.~Rozp{\k{e}}dek, M.~Pompili, A.~Stolk, P.~Pawe{\l}czak, R.~Knegjens, J.~de~Oliveira~Filho, {\em et~al.}, ``A link layer protocol for quantum networks,'' in {\em Proceedings of the ACM special interest group on data communication}, pp.~159--173, 2019.

\bibitem{di22}
S.~DiAdamo, B.~Qi, G.~Miller, R.~Kompella, and A.~Shabani, ``Packet switching in quantum networks: A path to the quantum internet,'' {\em Physical Review Research}, vol.~4, no.~4, p.~043064, 2022.

\bibitem{yo24}
S.~B. Yoo, S.~K. Singh, M.~B. On, G.~G{\"u}l, G.~S. Kanter, R.~Proietti, and P.~Kumar, ``Quantum wrapper networking,'' {\em IEEE Communications Magazine}, vol.~62, no.~3, pp.~76--81, 2024.

\bibitem{ca202}
M.~Caleffi and A.~S. Cacciapuoti, ``Quantum switch for the quantum internet: Noiseless communications through noisy channels,'' {\em IEEE Journal on Selected Areas in Communications}, vol.~38, no.~3, pp.~575--588, 2020.

\bibitem{qi99}
C.~Qiao and M.~Yoo, ``Optical burst switching (OBS)--a new paradigm for an optical internet,'' {\em Journal of High Speed Networks}, vol.~8, no.~1, pp.~69--84, 1999.

\bibitem{ma07}
A.~Martinez, J.~Aracil, and J.~L. De~Vergara, ``Optimizing offset times in optical burst switching networks with variable burst control packets sojourn times,'' {\em Optical Switching and Networking}, vol.~4, no.~3-4, pp.~189--199, 2007.

\bibitem{bh20}
M.~K. Bhaskar, R.~Riedinger, B.~Machielse, D.~S. Levonian, C.~T. Nguyen, E.~N. Knall, H.~Park, D.~Englund, M.~Lon{\v{c}}ar, D.~D. Sukachev, {\em et~al.}, ``Experimental demonstration of memory-enhanced quantum communication,'' {\em Nature}, vol.~580, no.~7801, pp.~60--64, 2020.

\bibitem{az23}
K.~Azuma, S.~E. Economou, D.~Elkouss, P.~Hilaire, L.~Jiang, H.-K. Lo, and I.~Tzitrin, ``Quantum repeaters: From quantum networks to the quantum internet,'' {\em Reviews of Modern Physics}, vol.~95, no.~4, p.~045006, 2023.

\bibitem{mu14}
S.~Muralidharan, J.~Kim, N.~L{\"u}tkenhaus, M.~D. Lukin, and L.~Jiang, ``Ultrafast and fault-tolerant quantum communication across long distances,'' {\em Physical Review Letters}, vol.~112, no.~25, p.~250501, 2014.

\bibitem{ni23}
D.~Niu, Y.~Zhang, A.~Shabani, and H.~Shapourian, ``All-photonic one-way quantum repeaters with measurement-based error correction,'' {\em npj Quantum Information}, vol.~9, no.~1, p.~106, 2023.

\bibitem{sh95}
P.~W. Shor, ``Scheme for reducing decoherence in quantum computer memory,'' {\em Physical Review A}, vol.~52, no.~4, p.~R2493, 1995.

\bibitem{st96}
A.~M. Steane, ``Error correcting codes in quantum theory,'' {\em Physical Review Letters}, vol.~77, no.~5, p.~793, 1996.

\bibitem{gt97}
D.~Gottesman, ``Stabilizer codes and quantum error correction,'' Ph.D. dissertation, California Institute of Technology, 1997.

\bibitem{mu16}
S.~Muralidharan, L.~Li, J.~Kim, N.~L{\"u}tkenhaus, M.~D. Lukin, and L.~Jiang, ``Optimal architectures for long distance quantum communication,'' {\em Scientific Reports}, vol.~6, no.~1, p.~20463, 2016.

\bibitem{so25}
I.~Sohn, B.~Kim, K.~Bae, W.~Song, C.~Lee, K.~Jeong, and W.~Lee, ``Fault-tolerant and secure long-distance quantum communication via uncorrectable-error-injection,'' {\em EPJ Quantum Technology}, vol.~12, no.~1, pp.~1--17, 2025.

\bibitem{ch20}
M.~Chiani, A.~Conti, and M.~Z. Win, ``Piggybacking on quantum streams,'' {\em Physical Review A}, vol.~102, no.~1, p.~012410, 2020.

\bibitem{rs01}
E.~Rosen, A.~Viswanathan, and R.~Callon, ``Multiprotocol label switching architecture,'' RFC 3031, 2001.

\bibitem{ie22}
IEEE, ``IEEE Standard for Ethernet,'' IEEE Std 802.3-2022, 2022.

\bibitem{co21}
T.~Coopmans, R.~Knegjens, A.~Dahlberg, D.~Maier, L.~Nijsten, J.~de~Oliveira~Filho, M.~Papendrecht, J.~Rabbie, F.~Rozp{\k{e}}dek, M.~Skrzypczyk, {\em et~al.}, ``Netsquid, a network simulator for quantum information using discrete events,'' {\em Communications Physics}, vol.~4, no.~1, p.~164, 2021.

\bibitem{wu21}
X.~Wu, A.~Kolar, J.~Chung, D.~Jin, T.~Zhong, R.~Kettimuthu, and M.~Suchara, ``Sequence: a customizable discrete-event simulator of quantum networks,'' {\em Quantum Science \& Technology}, vol.~6, no.~4, p.~045027, 2021.

\end{thebibliography}
\end{document}